# ElasticPlay: Interactive Video Summarization with Dynamic Time Budgets

Haojian Jin
Carnegie Mellon University
haojian@cs.cmu.edu

Yale Song
Yahoo Research
yalesong@yahoo-inc.com

Koji Yatani
The University of Tokyo
koji@iis-lab.org

## ABSTRACT

Video consumption is being shifted from sit-and-watch to selective skimming. Existing video player interfaces, however, only provide indirect manipulation to support this emerging behavior. Video summarization alleviates this issue to some extent, shortening a video based on the desired length of a summary as an input variable. But an optimal length of a summarized video is often not available in advance. Moreover, the user cannot edit the summary once it is produced, limiting its practical applications. We argue that video summarization should be an interactive, mixed-initiative process in which users have control over the summarization procedure while algorithms help users achieve their goal via video understanding. In this paper, we introduce ElasticPlay, a mixed-initiative approach that combines an advanced video summarization technique with *direct* interface manipulation to help users control the video summarization process. Users can specify a time budget for the remaining content while watching a video; our system then immediately updates the playback plan using our proposed cut-and-forward algorithm, determining which parts to skip or to fast-forward. This interactive process allows users to fine-tune the summarization result with immediate feedback. We show that our system outperforms existing video summarization techniques on the TVSum50 dataset. We also report two lab studies (22 participants) and a Mechanical Turk deployment study (60 participants), and show that the participants responded favorably to ElasticPlay.

## 1 INTRODUCTION

A recent investigation shows US adults spend 5.5 hours with video content (TV and online videos) per day [5]. However, this significant time allocation is still insufficient to explore the content users are interested in [1, 3]. Video engagement analysis shows the average watch time of a single Internet video is 2.7 minutes, which is much shorter than the average video length [1]. To perform quick content exploration, users actively use timeline widgets [32, 40, 54] to skip some footage.

To improve the demanding timeline manipulation for video browsing, previous work developed methods to fully automate the skipping process [25, 48]. However, two key problems still persist: inferring user needs and preferences [11], and balancing costs and benefits in skipping certain content [19]. We address these issues through a mixed-initiative approach [26], which takes advantage of the power of direct manipulation and automated summarization. The key challenge is to identify a control mechanism that is both intuitive for end users and practical for automated reasoning [12].

Our system incorporates a *time budget* control into a player interface, to allow users to specify intended time to watch the remaining part of a video. Time budget captures a real-world concept where users have different cost expectations for the remaining content based on their context (how busy they are at the moment) and the quality of content consumed so far. This notion of dynamic time budget is conceptually different from previous work that used content temporal constraints [7, 8, 29, 37, 46]. For example, the adjustable zoom factor [7, 8] lets users control how much detail a summary would contain, rather than how long the user will spend on the summary. Therefore, ElasticPlay goes further in the "user-centered" video interface direction, while the zooming metaphor is more "media-centered."

ElasticPlay packages time budget manipulation in a lightweight slider widget, as shown in Figure 1. Users can interact with the time budget in a similar manner to the traditional timeline widget. Once a user specifies a new time budget for the remaining content, ElasticPlay will trigger the shortening process and immediately adjust the future playback sequence to fit the new budget, providing direct, instantaneous feedback to the user.

ElasticPlay combines two commonly-used ideas to fit the remaining video content into a time budget: salient segment selection and fast-forwarding. Our system analyzes both audio and visual content to identify important parts and use this information to determine which parts to include, skip, or fast-forward. We leverage findings on content comprehension [15], and formulate shot selection and playback speed control into an optimization problem. Our system solves this optimization problem to maximize expected user comprehension under a time budget.

ElasticPlay is different from most previous work on video summarization in that it enables *interactive* video summarization. The primary goal of this work is to develop a human-centric intelligent video player that allows users to fine-tune the length of a summary *as they watch a video*, getting instantaneous feedback on how the summary would look like under different time budgets. Another notable aspect of our work is integration of fast-forwarding into video summarization. Most previous video summarization approaches produce a summary by either keeping or discarding parts of a video, often solving the standard 0/1 knapsack problem [48]. Instead, we formulate a variant of the 0/1 knapsack problem that finds the optimal playback speed of each part to maximize the content comprehension rate, based on the human comprehension model from the psychology literature [15, 34]. This allows us to naturally fast-forward parts that are somewhat less important, to the speed where users can still understand the content.

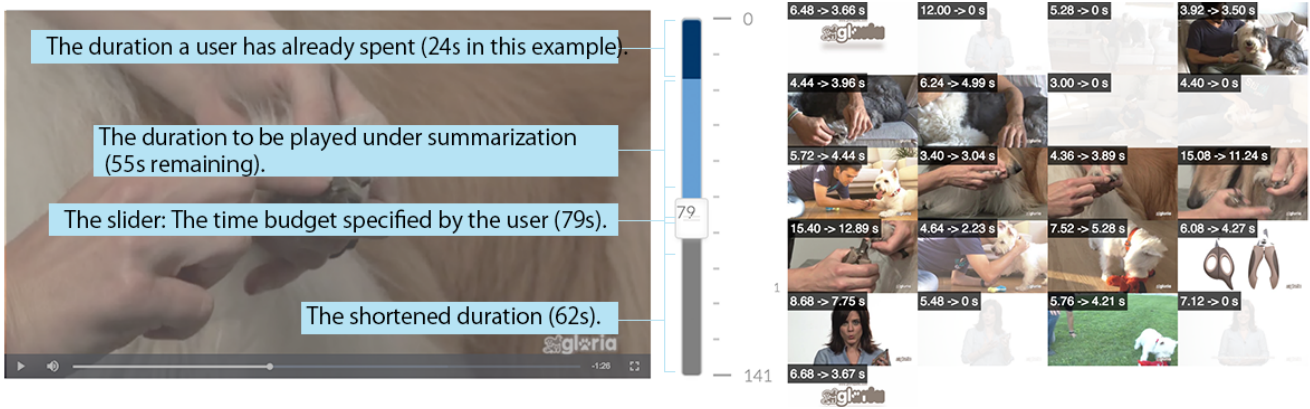

Figure 1: ElasticPlay is a video player widget that helps users watch a video with a time constraint. In this example, the user sets the time budget at 79s for a 141s video. The user has already spent 24s on the video and will need 55s more to finish the remaining 86s content. The right shot-by-shot visualization offers an overview of the playback strategy through the thumbnail grid. The transparency of each thumbnail indicates the percentage of skipped content in the corresponding shot. The numbers on each thumbnail show the duration change for each shot, e.g. 6.48->3.66 s means that ElasticPlay shortens a 6.48s shot to 3.66s through selective fast forwarding; 12.00->0 s means that ElasticPlay will skip that shot.

Our contributions in this work include:
- ElasticPlay, a mixed-initiative video interface incorporating the concept of time budget.
- The Cut-and-Forward (CaF) algorithm, a novel video summarization method with a combination of salient segment selection and selective fast-forwarding, formulated as an optimization problem with a variant of knapsack algorithm.
- Quantitative and qualitative evaluation on ElasticPlay, suggesting the benefits of increased transparency, live tuning and the hybrid shortening approach.

## 2 RELATED WORK

ElasticPlay is a *mixed-initiative* interface that builds upon previous work in *video direct manipulation* and *summarization*. Our *hybrid shortening* algorithm utilizes findings in user comprehension. Here, we review relevant work in these four related areas.

### 2.1 Mixed-initiative Video Interfaces

The idea of mixed-initiative interfaces was developed in 1990s [47]. Horvitz argued that the future interface research should focus more on exploring new direct manipulation or on automation via intelligent interactive agents [26]. Mixed-initiative interfaces was then proposed to leverage the complementary strengths of human and machine reasoning via a mixture of direct manipulation and interface agents [26, 38].

Previous work has attempted to involve users in the video browsing or summarization process [19, 35, 52], often referred as a "human-in-the-loop" process. For example, Tseng *et al.* [52] asked users to specify their preferences to visual concepts, such as sky, snow, a car, and a flag. Their system then determines whether to include particular shots based on the visual similarity to the preferred visual concepts. Adam *et al.* [7] allows users to dynamically filter video through specifying thresholds that determine the level of content granularity.

### 2.2 Video Summarization (Machine Reasoning)

Video Summarization, in which a system produces a shortened version of a video, is generally intended to support content skimming and searching in a large video corpus [25]. Researchers have developed various algorithms to generate an efficient summary (e.g., based on visual attention [13, 18], content frequency [21, 56] and non-redundancy [22, 36, 55]).

While content analysis technologies have improved significantly in the past years, most summarization applications are still for screening purposes, such as surveillance video search [10, 24], sports video highlighting [16], and preview generation [23, 48]. These applications help users determine whether to watch the video in full-length [53]. ElasticPlay applies summarization methods to creating new user experience for video browsing.

### 2.3 Video Navigation (Direct Manipulation)

Research on video navigation techniques explores new ways to allow users to access video content. For example, through extracting motion trace, Dragicevic *et al.* [17] enabled users to control video playback by dragging objects of interest along their trajectory. By visualizing users collective interaction peaks in the timeline widget, Kim *et al.* [31] informed users of the importance of each segment in MOOCs videos.

Researchers also looked at external data sources to explore new ways to index video content. Pavel *et al.* built systems for video search and browsing using a crowdsourcing mechanism [43]. Their follow-up system further incorporated synchronized captions, scripts, and plot summaries [42]. In their systems, users can find relevant clips with expressive search at three levels: dialogue, shot-by-shot actions, and high-level plot events.

These interfaces and systems offer different content indexing techniques, which help users manipulate the timeline widget and save time indirectly. ElasticPlay differs from them in allowing users to directly control a summarization process to accommodate their time-saving needs.

## 2.4 Fast-forwarding

Recent research prototypes [24, 28, 30, 34, 44] and production systems (e.g., Youtube, edX and Overcast) started to incorporate the fast-forwarding feature. For example, SmartPlayer [16] dynamically adjusts playback speeds based on the complexity of current scene and predefined semantic events. Pongnumkul et al. [46] developed a content-aware dynamic timeline control to navigate long videos.

Although fast-forwarding is easy to implement and effective in time-saving, previous research shows that users' comprehension rate reduces as the playback rate increases [30, 34, 49]. The perception of audio/video fast-forwarding follows the Weber-Fechner law (WF-law), which predicts that sensation is proportional to the linearity/logarithm of the stimulus intensity under/above a certain threshold [15, 20, 30, 34, 49]. ElasticPlay takes this theory to build a comprehension model for continuous playback rate adjustment.

## 3 FORMATIVE STUDY

To obtain interface design insights, we conducted a series of formative studies to compare the following five methods commonly used in video browsing.
(1) *Manual skipping*: Users could skip any portion of video by using a timeline widget embedded in the standard video player.
(2) *Manual control on the playback speed*: Users could control the playback speed by pressing "up" or "down" arrow keys on the keyboard.
(3) *Constant fast-forwarding*: The system fast-forwarded the entire video with a constant speed.
(4) *Adaptive playback speed (audio-based)*: The system determined an adaptive playback rate based on the detected beat-per-minute (BPM) in audio streams [9, 16].
(5) *Adaptive playback speed (motion-based)*: The system determined an adaptive playback rate based on the number of motion vector changes per minute in video streams [45].

We recruited 10 participants (6 male, mean age 27.2, max=33, min=22), who were full-time employees in an IT company, and asked them to watch videos with given interfaces under two time-limited settings: 33% and 66% of the original video length. Our qualitative results revealed the design trade-off for each method.

**Manual methods: Excessively demanding.** Our participants found manual methods useful but mentally demanding. In particular, they struggled to determine which portion of the video to skip. Most participants were dissatisfied with this experience, and not able to achieve an ideal time allocation.

*"I cannot find where to skip. If someone will soon start speaking, I would not jump. Most of the time, I just jumped to the middle of a speech, and I needed to go back-and-forth many times to get to the right moment. It takes too much effort to do this but helps a little for the understanding. The manual control is horrible."*

**Constant fast-forwarding: Limited shortening capability.** Participants mentioned that they were not able to understand videos at 3X playback (i.e. 33% of the original video length). This result is in line with previous findings [34, 49].

*"I can understand most videos at the 66% budget, but I cannot learn anything from the audio at the 33% budget – My understanding mainly relies on visual content. If I watch videos on a platform with speed options, I usually set the speed at 1.25x, or 1.5x if the video does not contain much speech content."*

**Adaptive playback speed: speech and semantic distortion.** Our participants commonly identified that playback speed did not seem consistent and reported undesired semantic distortions. Existing fast-forwarding systems usually mute the audio to avoid this issue [16].

*"The video looks weird. The speed changes all the time during the speech. I have no clue what the playback rate would be and if it associates with some intent."*

The results led us to derive the following design implications.
- **The system should combine segment selection and fast-forwarding.** Our results revealed benefits of segment selection and fast-forwarding though both need substantial improvements. Segment selection can shorten the video significantly, but may break the flow of a story. While fast-forwarding can keep it, the shortening capability is limited. Thus, our system should exploit these benefits to support improved video watching under a time budget.
- **The shortening process should be consistent and transparent to users.** Participants expressed concerns about the black box process of adaptive playback and intelligent skipping. Making the shortening process transparent can help users build a confidence of the automatic skipping.
- **The system should allow users to set their time budgets.** We observed participants' distinct time allocation preferences for the same videos. Moreover, we also found that the participants often wanted to change the time budget even during the course of video watching. Thus, the time budget control should be interactive.

## 4 ELASTICPLAY

The findings from our formative studies clearly suggest that an ability to set a dynamic user-defined time budget would improve video watching experience. This leads us to design a mixed-initiative interface for video summarization.

### 4.1 System Overview

ElasticPlay comprises three main modules: the offline frame analysis module (on a server side); the online playback strategy decision module (on a user's browser side); and the user interface. The overall process of finding an optimal playback strategy is as follows (Fig 2):
(1) **Input file augmentation**. ElasticPlay first collect still images from online image search by using the title of the given video.
(2) **Offline frame analysis**. ElasticPlay then performs visual content analysis by measuring the visual similarity between video content and search results. It also executes voice activity detection. This analysis produces shot importance scores and speech annotation.
(3) **Online playback strategy generation**. When the video is initially loaded in the browser, ElasticPlay will construct a solution cache for different playback speeds and time budgets based on the pre-processing results.

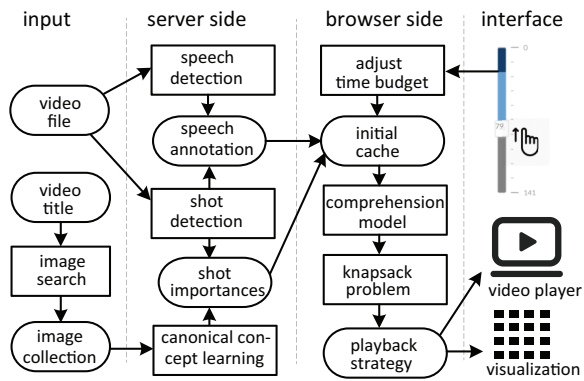

Figure 2: An ElasticPlay system architecture overview.

(4) **Upon user interaction with ElasticPlay**. Based on the time budget specified by a user, ElasticPlay searches the optimal playback plan in the constructed solution cache.

Pre-processing (step 1, 2) needs to be performed only once, ideally as soon as new videos become available in the data server; hence it would not affect the real-time user experience.

We implement both the interface and the decision module as a lightweight JavaScript library through HTML5 APIs [4, 6]. Developers can incorporate the ElasticPlay widget into a standard HTML5 video player with less than 20 lines of code[1].

### 4.2 Interface

Figure 1 shows our ElasticPlay interface. The entire length of the slider represents the amount of time a user will spend watching the video in full-length. The dark blue area (top) represents the time a user has already spent watching the video while the light blue area (middle) represents the remaining time. The gray portion represents time saved by the summarization. The slider handle is initially placed at the full-length position (bottom). Users can specify time budget by simply moving this handle at any time during the video is playing. Any change on the location of the slider handle triggers dynamical updates on the playback plan, and ElasticPlay immediately applies it to the current video watching.

ElasticPlay offers an additional visualization tool (Figure 1c), which provides a grid of consecutive shots. This is updated in real-time as the slider location changes. The two numbers at the left-top corner of each thumbnail are the original and shortened length in the current playback plan, respectively. This visualization is intended to provide additional transparency in the process, and it is hidden to users by default.

## 5 CUT-AND-FORWARD ALGORITHM

The core component of ElasticPlay is our novel Cut-and-Forward (CaF) algorithm, which integrates salient segment selection and selective fast-forwarding. Given a video, CaF performs video segmentation and computes shot importance scores, using speech detection and visual analysis (details below). The algorithm then solves a variant of the Knapsack problem to find an optimal video summarization strategy, by removing or fast-forwarding some of less important shots. A high-level explanation of our algorithm is as follows. We first fast-forward the speech and non-speech content at different rates. This is because fast-forwarding non-speech content is less noticeable to users. We then drop segments if comprehensible speeds are not available at a given time budget. We describe the details of our CaF algorithm in this section.

[1] https://github.com/haojian/elasticplay

### 5.1 Video Segmentation and Analysis

The frame analysis module first runs voice activity detection (VAD) [2] to distinguish content with and without speech. It also executes a shot detection algorithm [41] to divide the videos into shots based on the visual differences. Intersecting these two results, we annotate the speech and non-speech content in each shot.

We then measure the importance score of each shot using the title-based video analysis method developed by Song et al. [48]. The method assumes that a title of a video is chosen to be maximally descriptive of its main topic; hence, images related to the title can serve as a proxy for important visual concepts in a video. The method first collects still images by using a title as a web search query. It then evaluates the importance score of each frame in a shot by measuring its visual similarity to the set of collected images using a variant of sparse coding. We calculate the shot importance score as the average across all frames in each shot. The process takes about two minutes for a five-minute video on a machine with a 2.3 GHz Intel Core i7 CPU.

### 5.2 Estimated Comprehension Model

To formulate our video shortening as an optimization problem, we use a human comprehension model on videos. Psychology studies [15, 20, 34, 49] show that the comprehension rate decreases as the playback speed increases. We compute apparent importance ($AI$) of each shot by multiplying the comprehension rate $p$ and importance score $s$ of each shot. We then estimate the comprehension of the whole video by summing $AI$ across all shots:

$$AI_{all} = \sum_i p_i s_i \quad (1)$$

where $p_i$ is the comprehension rate of the $i$-th shot, and $s_i$ is the importance score.

Existing research [15, 34] also studies the relationship between playback speed $v$ and the comprehension rate $p$ (a relative scale). It reveals that the relationship can be described as a linear model if the playback rate is under certain thresholds [49]. This leads us to estimate the comprehension rate $p \in [0, 1]$ as:

$$p = k \times (v - 1) + 1, \quad where \quad 1 \leq v \leq VT \quad (2)$$

where $k$ is the WF-law constant (a negative value) and $VT$ is the playback speed threshold. This model assumes that viewers would have perfect comprehension when a video plays in a normal speed (i.e., $v = 1$).

### 5.3 Selective Fast-forwarding

Fast-forwarding reduces the video comprehension rate because users cannot fully understand the content if the playback speed is above a certain threshold. Kurihara et al. [34] showed that the playback speed threshold and WF-law constant are different for content with and without speech. This motivates us to develop selective fast-forwarding, which speeds up the content with speech and without speech separately. Based on the data reported by [15, 34], we set:

$$VT^n = 6, VT^s = 1.5, k^n = -0.1, k^s = -1 \quad (3)$$

where $VT^n$ and $k^n$ are the threshold and WF-law constant for non-speech content, and $VT^s$ and $k^s$ are those for speech content.

After selective fast-forwarding, we have the apparent importance $\widetilde{s_k}$ for $k$-th shot:

$$\widetilde{s_k} = \widetilde{s_k^n} + \widetilde{s_k^s} = s_k^n(k^n \times (v^n - 1) + 1) + s_s^n(k^s \times (v^s - 1) + 1) \quad (4)$$

where $v^s$ and $v^n$ are the playback speeds and $s_k^s$ and $s_k^n$ are the importance scores for shots with and without speech, respectively. The length of the shortened shot $\widetilde{l_k}$ is:

$$\widetilde{l_k} = \frac{l_k^n}{v^n} + \frac{l_k^s}{v^s} \quad (5)$$

## 5.4 A Variant of the Knapsack Problem

Using Equation 1 and 2, CaF finds an optimal combination of fast-forwarding and content skipping given a time budget. We formulate this problem as a variant of 0/1 knapsack problem [48].

The traditional 0/1 knapsack problem is formulated as follows. Let U = { $u_1, u_2, \cdots, u_n$ } denote a set of variables indicating whether we include or exclude each of video shots. Each $u_i$ can take a binary value (0 or 1), and $u_i = 1$ when the shot of $i$ is included. Let S = { $s_1, s_2, \cdots, s_n$ } denote the importance scores of the shots. In order to generate a video of length under a budget $B$, the 0/1 knapsack problem solves the following optimization problem:

$$max \sum_{i=1}^{n} u_i s_i, \quad subject\ to \quad \sum_{i=1}^{n} u_i l_i \leq B, \quad u_i \in \{0, 1\} \quad (6)$$

Unfortunately, the traditional setting does not allow us to incorporate the idea of fast-forwarding, because the variables in $U$ are binary. To enable selective fast-forwarding, we replace the importance scores $s_i$ with our apparent importance scores $\tilde{s}_i$:

$$max \sum_{i=1}^{n} u_i \widetilde{s_i}, \quad subject\ to \quad \sum_{i=1}^{n} u_i \widetilde{l_i} \leq B, \quad u_i \in \{0, 1\} \quad (7)$$

To solve this variant of knapsack problem, we discretize both the speed ranges $[1, VT^n]$ and $[1, VT^s]$ into N steps, resulting in $N^2$ speed combinations. This leads to $N^2$ standard knapsack problems. ElasticPlay then solves each knapsack problem through dynamic programming and keeps the computed results in a cache to support future interactions. In practice, this pre-computation takes less than a second even for videos longer than 10 minutes.

We set $N$ to 10 in the current implementation, which provides speed granularity of 0.05x for content with speech and 0.5x for content without speech. The CaF algorithm runs the brute force search when the user specifies a time budget, and selects a solution that has the overall highest score. The resulting video is a concatenation of shots with $u_i = 1$ in the chronological order.

## 5.5 Reusable Knapsack Cache for Interactivity

The time complexity for a standard knapsack problem is $O(n \times W)$, where $n$ is the number of shots and $W$ is the knapsack capacity. For the traditional setting with a pre-determined time budget, $W$ equals to the time budget, either in seconds or in frames [48].

One major challenge in our CaF algorithm is to maintain responsiveness. A naïve solution is to solve our variant knapsack problem each time a user changes the time budget, which has the same time complexity as the standard setting. But this would hinder getting results in real-time, deteriorating responsiveness.

We illustrate this problem with an example. The key idea behind dynamic programming for a knapsack problem is to compute a matrix-like cache $M$ in a recursive manner. Each cell (denoted $M(i, j)$) stores the maximum importance score sum of a subset of $S_1, S_2, .., S_i$ given the duration of $j$ frames. With the standard frame rate (30 frames per second), $j/30$ yields the time to play. Suppose that a user initially sets the time budget to 100 seconds for a 200-second video containing 50 shots at 30 frames per second. When the video is loaded, an algorithm calculates the matrix of $M$. In this matrix cache, $i$ is between $1 \cdots 50$ and $j$ is $1 \cdots 200 * 30$, and thus the size of $M$ is $50 \times 6000$. Note that each cell $M(i, j)$ caches the optimal playback plan for the first $i$ shots at a time budget $j$.

However, the problem arises when the user changes the time budget while watching a video. Suppose that the user decides to increase the time budget from 100 seconds to 140 seconds, after watching 20 shots for 30 seconds. The algorithm now has to calculate an optimal solution from a subset of $S_{21}, S_{22}, .., S_{50}$. Unfortunately, the algorithm cannot reuse $M(i, j)$ at this point because the shot set becomes different. It needs to re-calculate $M$ with a different shot candidates pool.

To remedy this issue, we construct a cache about shots in the reverse order. Each cell $\overline{M(i, j)}$ in the new cache matrix stores the optimal playback plan that fills $j$-frame playback duration with a subset of shots $S_{n-i+1}, S_{n-i+2}, .., S_n$. In the example above, the CaF algorithm calculates the solution cache matrix $\overline{M}$ in the same manner. When the time budget is updated, the CaF algorithm looks for a solution with a subset of $S_{21}, S_{22}, .., S_{50}$ from $\overline{M(30, (140 - 30) \times 30)} = \overline{M(30, 3300)}$. It then performs dynamic programming from $\overline{M(30, 3300)}$ to reconstruct an optimal playback plan for the given time budget and remaining shots. The initial cache computation procedure takes $O(N^2 \times n \times W)$, where $N$ is different speed settings, and $W$ is the maximum time budget (the total length of the original video, in frames). The later optimal solution lookup takes only $O(n)$.

## 6 EVALUATION

We conducted a series of quantitative and qualitative studies to evaluate the ElasticPlay interface and CaF algorithm. We first validated the performance of the CaF algorithm through quantitative evaluations. We then evaluated the user experience of videos summarized through the CaF algorithm in a controlled study. Finally, we tested the benefits of interactivity through a laboratory and crowdsourcing study.

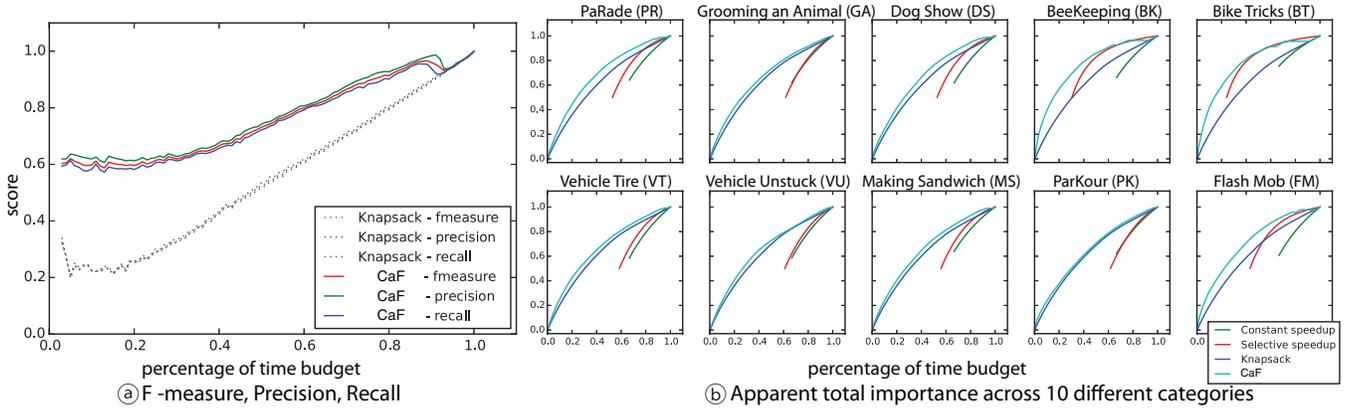

**Figure 3: Overall the CaF algorithm can provide better content coverage and expected content comprehension across different time budgets. However, the specific performances were correlated with the video types.**

## 6.1 Quantitative Algorithm Evaluation

*6.1.1 Content Coverage.* We first evaluated content coverage of our shortening method, a common metric in the video summarization literature [27, 48]. The novelty of our CaF algorithm lies in the integration of fast-forwarding, which is independent from the video analysis. Thus, instead of measuring the absolute content coverage, we evaluated the relative performance gain from the integration of fast-forwarding.

We compared our CaF algorithm with the commonly used knapsack method with the TVSum50 dataset [48]. Both the CaF algorithm and naïve knapsack method used the same computed importance scores as we explained. The ground truth data were the human annotated importance score with each summarization method. We then measured agreements between the resulting summary and the ground truth.

As each video contains ratings by 20 people, we had 20 different ground truth annotations. We computed precision, recall and $F_1$-measure between each pair of a generated summary and ground truth, and derived the average value. This average pairwise $F_1$-measure is as follows:

$$F_1 = \frac{1}{20} \sum_{i=1}^{20} \frac{2(p_i \times r_i)}{p_i + r_i} \quad (8)$$

where $p_i$ and $r_i$ are the precision and recall, respectively.

Figure 3a shows the average pairwise $F_1$-measure, precision and recall across different time budgets for all 50 videos. When the time budget was sufficient, salient shot selection dominated the shortening process. Thus, the naïve knapsack method had a similar performance with CaF. Once fast-forwarding started to be involved, CaF outperforms the knapsack method. This is because, as shots are fast-forwarded, the time budget in CaF is equivalent to a larger time budget in the naïve knapsack method. Overall, the CaF algorithm can provide contents of better relevance (recall) and higher quality (precision) at nearly all the different time budgets.

*6.1.2 Expected Content Comprehension.* The techniques we used in the CaF algorithm can be broken down into three elementary techniques: constant fast-forwarding, selective fast-forwarding, and naïve shot selection (i.e., naïve knapsack). We examined content comprehension rates against these techniques.

Figure 3b plots content comprehension rates of the four methods including CaF across different categories (categories code can be found in [48]). We normalized the scores so that they were 1.0 for the 100% time budget (i.e., no shortening). For the fast-forwarding, we tested only up to the threshold described above because otherwise videos became incomprehensible.

We found that the performances of these four different methods were largely correlated with the speech portion of videos. In most categories, the CaF and naïve knapsack methods demonstrated the best performance. The selective fast-forwarding method was the second best methods in three categories (BK, BT, and FM). Overall, the CaF algorithm performed better than the other three methods in the majority of cases across the categories and time budgets.

## 6.2 User Experience of CaF-generated Videos

*6.2.1 User Study Design & Procedure.* We prepared three conditions: 1) Constant fast-forwarding (*FF*); 2) the 0/1 Knapsack method [48] (*KS*); and 3) the CaF algorithm. We chose two videos from the TVSum50 dataset [48] for this user study. Both videos were instructional videos: "tire fixing" and "making a sandwich." We chose these two videos as they were relatively easy to identify content loss in shortened versions. We also set two different time budgets of 33% and 66% of the original duration. The value of 66% was chosen because we expected that fast-forwarding would still show acceptable performance. The other value was chosen to represent a more time-constrained case. As a result, we had six summarized versions for each video.

Our study consisted of two sessions, each of which involved either of the videos. At the beginning of a session, participants were first asked to view the original video. They kept watching the video until they felt that they had fully understood the content. We then presented the six shortened videos to the participants one-by-one. Every time they completed watching, the participants answered their impression about how comfortable the watching experience is in the 5-Likert scale. We randomized the orders of video presentation to minimize any possible bias.

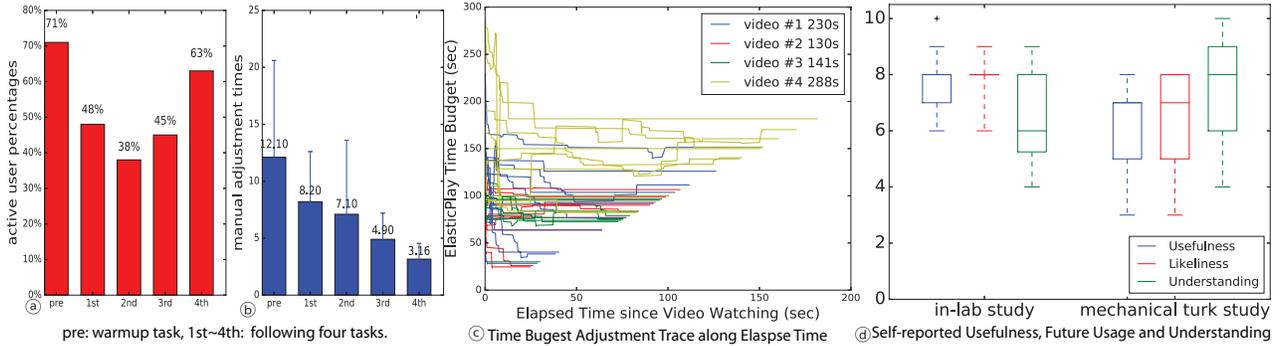

Figure 4: Left: Participants are willing to use ElasticPlay voluntarily. Middle: Participants can build the mental model of ElasticPlay quickly. Right: Participants feel positive in the self-reported survey.

We recruited 12 participants (7 male, mean age 29.7, max=49, min=21). Six of them were native English speakers, and the rest lived in English speaking countries for more than five years.

*6.2.2 Results.* All the interfaces received a comfortable rating at 66% time budget (CaF: $\mu$ = 4.06, $\sigma$ = 1.00 , FF: $\mu$ = 4.11, $\sigma$ = 0.83, KS: $\mu$ = 3.89, $\sigma$ = 0.90). However, the rating of Fast-forwarding reduced significantly at 33% time budget (CaF: $\mu$ = 3.06, $\sigma$ = 1.06 , FF: $\mu$ = 1.89, $\sigma$ = 0.96, KS: $\mu$ = 2.94, $\sigma$ = 1.00). Our Mann-Whitney test confirmed that the CaF and knapsack methods were significantly better than the fast-forwarding approach ($p < .01$). We thus concluded that our CaF algorithm offered more stable and conformable user experience than the other two methods.

At the 33% condition, CaF and KS show a similar performance. This is, in part, because content skipping dominates the shortening process. Our videos were relatively short (3~4 mins), so fast forwarding used in CaF would include only a few more shots than KS. This made the two resulting videos similar to each other.

## 6.3 ElasticPlay User Experience

Our quantitative results revealed benefits of the CaF algorithm. To examine the experience of ElasticPlay as a whole, we conducted another user study in our laboratory (10 lab participants) as well as through Amazon Mechanical Turk (AMT; 60 participants). Our main objective of this part of the studies is to validate the design of ElasticPlay and confirm how easily people would adapt to responsive video browsing through time budgets.

*6.3.1 User Study Design.* We instructed participants that our main objective of this study is to understand how people would estimate an optimal time budget for a video. In this manner, we were able to examine how often our participants voluntarily used ElasticPlay for their video watching. We first provided participants with a one-page instruction about ElasticPlay, then presented five video watching tasks (one warm-up task + four tasks). To highlight the benefits of ElasticPlay's responsiveness, we asked participants to estimate an initial time budget based on the meta-data. The system then presented a summarized video with that initial time budget (Figure 1). The participants were allowed to change their time budget later in the course of video watching if they wished, but not enforced. We used five videos from the TVSum50 dataset and presented these videos in a randomized order. Our system recorded all interactions, including mouse movements, time budget adjustment, and text input for further analysis.

At the end of the study, we asked the participants three 10-Likert scale questions (1: most negative – 10: most positive): "how useful was the slider widget for watching video?", "how likely would you use it in your real life?", and "how well do you think you understood the summarization process along different time budgets?".

To avoid malicious task completion [33], we asked AMT participants to summarize the video into a short paragraph (more than 30 words) after watching. We also required >95% HIT approved rates before participation.

*6.3.2 Participants.* We recruited 10 lab participants (4 male, mean age 22.9, max=25, min=21) in the in-person setting. All participants are different from the earlier studies and have no prior knowledge about ElasticPlay.

In AMT, we released 60 HIT tasks and paid $3 upon completion. 83 AMT workers participated, but only 60 completed the whole study. We excluded 10 participants who spent less than 10 seconds on the instruction page. These participants either did not use the slider widget at all during the study or spent less than five minutes on the whole study. The remaining 50 MT participants spent 8 to 29 minutes to complete the study ($\mu$ = 16.31 mins, $\sigma$ = 6.79) with allocating 65.23s ($\sigma$ = 25.87s) on the instruction page on average. They also provided sufficiently long text descriptions (# of words in the text: $\mu$ = 57.38, $\sigma$ = 20.97). We determined to include all these 50 AMT participants for our analysis.

*6.3.3 Results.* We examined the occurrences of interaction on the time budget widget, with two hypotheses: 1) our participants would continue to use the widget several times because they realized the benefits of ElasticPlay, and 2) our participants would interact with it less frequently as they build the mental model of ElasticPlay. We used two metrics to verify our hypotheses: Active User Percentages (AUP = [the number of participants who used the slider] / [the total number of participants]) and average Manual Adjustment Times (MAT = [the number of slider adjustment interaction in each task]).

Figure 4a illustrates the active user percentages across sequential tasks in both studies. "Pre," "1st," "2nd," "3rd," and "4th" represent the warm-up and following four tasks, respectively. At the warm-up task, 71% of our participants used the time budget widget. During the experiment, AUP ranged between 38% and 63%. Although the AUP values varied a little across the session (Fig 4a), they imply that our participants were relatively willing to keep using ElasticPlay.

Figure 4b further shows the average MAT across the warm-up and following four tasks. We considered a time budget change as an independent event if it occurred more than 500ms after the previous widget manipulation. As shown in Figure 4b, the average MAT exhibited a decreasing trend over the tasks. The decreasing MAT (Fig 4b) suggests that our participants were able to become efficient to achieve perceived optimal playback strategies over time. This implies that our participants successfully developed a correct mental model of ElasticPlay.

Figure 4c illustrates the actual traits of time budgets specified by our lab participants. The x and y-axes represent the time the lab participants spent on the video and the time budget specified at the given moment, respectively. Since ElasticPlay always finishes videos at a specified time budget, each trait ends before their actual length. The raw trace plot suggests that most lab participants tended to conservatively estimate time budgets and gradually tuned/reduced them during the course of video watching. One of the lab participants described this experience as follows:

*"I don't want to miss any important scene in the video. So my strategy is setting a large initial value to start the video. I usually reduce the budget when the content is boring, and the next shots seem in the same scene."*

The post study self-reported score were also positive. Both Lab and MT Participants agreed that ElasticPlay was useful to complement their video watching interaction (Lab: $\mu$ = 7.8, $\sigma$ = 1.08; MT: $\mu$ = 6.13, $\sigma$ = 2.09), they would use it in the future (Lab: $\mu$ = 7.9, $\sigma$ = 0.83; MT: $\mu$ = 6.27, $\sigma$ = 1.73) and the interface helped them understand the summarization process better (Lab: $\mu$ = 6.6, $\sigma$ = 1.68; MT: $\mu$ = 7.73, $\sigma$ = 1.77) (Fig 4d).

In the post-study interview, lab participants also agreed that the shot visualization helped them understand the shortening process. Two participants expressed that they wanted the visualization to be shown as an overlay on a video when the slider is operating, but to be hidden otherwise.

## 7 DISCUSSION AND LIMITATIONS

**ElasticPlay as an Interface**: The contributions of our interactive summarization and hybrid shortening algorithm are independent of specific video analysis methods. Our ElasticPlay implementation does not require any additional data except the video and its title. In the future, ElasticPlay can be applied to other analysis methods, such as collective behavior patterns [31], hierarchical analysis [39, 55], crowd labeled data [16], and movie scripts/subtitles [42, 43].

**Fast-forward v.s. Intelligent skipping**: Despite the recent developments of auto-skipping with advanced video summarization methods, constant fast-forwarding is still the dominant video skimming technique. During our study, participants expressed their skeptical attitude toward auto-skipping technologies.

*"I gave the rating based on several checkpoints. While I have seen all the important points in the summary, I am still afraid of missing some content."*

Fig 4c also suggested participants' conservative estimation on the time budget. ElasticPlay can alleviate this problem as it provides direct and flexible control on summarization through a time budget.

**Comprehension Model**: We do not claim that the WF-law modeling is the optimal way to model user comprehension. Exploring the relationship between playback speeds and understandability has been an active topic in psychology research [14, 20]. We note that ElasticPlay is independent of a particular comprehension model. ElasticPlay can thus potentially build on other comprehension models in the future.

Users may have different preferences over fast-forwarding and selective shortening. For example, the lower upper limit of the playback rate could be desirable for non-native speakers. By reducing the WF-constant for non-speech content, our method will weigh more on skipping. Future work should explore how system parameters can be determined for personalization.

**Enjoyment Model**: Video watching has an entertainment aspect. People watch videos not only for seeking particular information but also for sparing time or enjoying themselves [50]. Our current ElasticPlay implementation does not consider enjoyment perceived by users. Such a metric [51] could improve identification of important shots within a footage, potentially leading to smoother and better summarized videos.

**Video Types**: During the study, we found that the performance of the hybrid shortening method may vary depending on video content or style. Our study on the comprehension rate (Figure 3b) shows that the performance is relative to the speech portion. Future work should further examine the performance of our CaF algorithm with a wider variety of videos.

**Video length in user studies**: All of our user studies involved relatively short videos. A deployment study may help to understand how similarly or differently people would use ElasticPlay for watching long videos in comparison to our findings.

## 8 CONCLUSION AND FUTURE WORK

As the amount of video content continues to grow, summarization is becoming a major method to help people browse videos. In this paper, we describe ElasticPlay, an interface widget that enables interactive video summarization. It allows users to drive video summarization through their time budgets. To maximize the summary comprehension, our CaF algorithm combines two shortening methods of selective fast-forwarding and selective shot selection for summarization. Our evaluations revealed improved summarization accuracy of CaF and smooth user adaptation to the ElaticPlay interface. Future work should examine the performance of ElasticPlay with more video quality metrics, such as visual continuity and enjoyment. Psychological implications by ElasticPlay are another interesting topic for future work.